\documentclass[showpacs, pra,twocolumn,preprintnumbers ,amsmath, amssymb, superscriptaddress, aps]{revtex4-2}

\usepackage[utf8]{inputenc}
\DeclareUnicodeCharacter{202F}{\,}
\usepackage{color}
\usepackage{amsmath,amssymb}
\usepackage{pifont}% http://ctan.org/pkg/pifont
\usepackage{amssymb}  % More symbols
\usepackage{bbold}
\usepackage{float}
\usepackage{subfloat}
\usepackage[squaren]{SIunits}
\usepackage{xcolor}

\usepackage[caption=false]{subfig}
\usepackage{tikz}
\usepackage{makecell}
\usepackage{subfig}
\usepackage{pifont}   % Ding symbols
\usepackage{graphicx} % Include figure files
\graphicspath{{Figures/}}
\usepackage{dcolumn}  % Align table columns on decimal point
\usepackage{bm}       % bold math
\usepackage{multirow} % Table functions
\usepackage{placeins}% For making floats not move around everywhere
\usepackage[colorlinks]{hyperref}
\usepackage{mathtools}
\usepackage{appendix}

\begin{document}
		\title{Tuning quantum tunneling in WSe$_2$ via strain engineering}
			%Strain effect on tunneling through a static barrier in WSe$_2$}
	
	\date{\today}
	
	\author{Rachid El Aitouni}
	\affiliation{Laboratory of Theoretical Physics, Faculty of Sciences, Choua\"ib Doukkali University, PO Box 20, 24000 El Jadida, Morocco}
	
	\author{Hasna Chnafa}
	\affiliation{Laboratory of Theoretical Physics, Faculty of Sciences, Choua\"ib Doukkali University, PO Box 20, 24000 El Jadida, Morocco}

	\author{Clarence Cortes}
	\affiliation{Vicerrector\'ia de Investigaci\'on y Postgrado, Universidad de La Serena, La Serena 1700000, Chile}  
	\author{David Laroze}
	\affiliation{Instituto de Alta Investigación, Universidad de Tarapacá, Casilla 7D, Arica, Chile}

	\author{Ahmed Jellal}
	\email{a.jellal@ucd.ac.ma}
	\affiliation{Laboratory of Theoretical Physics, Faculty of Sciences, Choua\"ib Doukkali University, PO Box 20, 24000 El Jadida, Morocco}
	%\affiliation{Canadian Quantum  Research Center, 204-3002 32 Ave Vernon,  BC V1T 2L7,  Canada}
		\begin{abstract}
	
% We investigate the effect of strain in the presence of a scalar potential on transmission, conductance, and the polarization of spin and valley, with the aim of enhancing its technological applications by tungsten diselenide WSe$_2$. The applied barrier divides the sheet into three regions. The wave functions are derived analytically, and their continuity at the barrier interfaces allows us to determine the transmission and reflection coefficients. The transmission probability is calculated using current densities. The Büttiker relation is then applied to evaluate the conductance, which in turn enables the calculation of spin and valley polarization. Numerical results demonstrate that the properties of WSe$_2$ are influenced by strain, the applied potential, and the incident energy. The transmission and conductance show pronounced oscillations resulting from quantum interference effects, while spin and valley polarizations exhibit significant and controllable variations. These findings indicate that effective control of electronic transport in WSe$_2$ may enable advanced spintronic, valleytronic, and optoelectronic applications.

We present a comprehensive theoretical study of strain-engineered quantum transport in monolayer tungsten diselenide (WSe$_2$) in the presence of an electrostatic scalar potential. By incorporating strain effects within a low-energy Dirac framework, we analyze their impact on spin- and valley-resolved transmission, conductance, and polarization. The applied potential barrier partitions the system into three distinct regions, allowing for an analytical derivation of the wave functions in each domain. Enforcing continuity conditions at the interfaces yields exact expressions for the transmission and reflection amplitudes. The transmission probability is evaluated from the corresponding current densities, while the conductance is obtained using the Landauer--B\"uttiker formalism, enabling a quantitative determination of spin and valley polarizations.
Our numerical analysis reveals that strain acts as a powerful tuning parameter that reshapes the electronic dispersion and strongly modifies transport characteristics. In particular, the transmission and conductance exhibit pronounced oscillatory behavior driven by quantum interference and resonant tunneling mechanisms. More importantly, both spin and valley polarizations display substantial and highly controllable variations as functions of strain, barrier height, and incident energy.
These results demonstrate that strain and electrostatic engineering provide an efficient and versatile platform for manipulating spin--valley degrees of freedom in WSe$_2$. The ability to tailor polarization and interference effects suggests promising opportunities for the design of next-generation spintronic, valleytronic, and optoelectronic devices based on two-dimensional transition-metal dichalcogenides.

	\end{abstract}
		\pacs{72.80.Vp, 73.23.-b, 78.67.-n\\
		{\sc Keywords:}
		%% keywords here, in the form: keyword \sep keyword
		Monolayer WSe$_2$, strain, scalar potential, Dirac equation,  transmission, Klein tunneling, conductance.}
	\maketitle
	\section{Introduction}	\label{Intro}

    The electronic and optical characteristics of two-dimensional (2D) materials stand out when you compare them with traditional semiconductors, which include silicon \cite{Deborde2024,Li2025,Zhao2023}. Their atomically thin structure enables them to achieve carrier mobilities which surpass the performance of ultra-thin silicon films \cite{Ahmed2017,2DMobilityReview2020}. 
The bandgap of various 2D materials can be easily modulated by quantum confinement \cite{Zhao2023,2DBandgapTuning2025}, and this in turn tailors the interaction of light with matter \cite{LightMatterTMDs2017,EnhancedOpticalTMDCs2022}, and allows multiple optoelectronic and nanoelectronic devices to be fabricated \cite{2DApplicationsReview2024,TMDDevices2017,2DOptoelectronics2025,MoS2NextGenReview2025}. Graphene represents the first 2D material ever isolated having a hexagonal structure with extraordinary electronic and thermal properties \cite{Novoselov2004,Geim2007,CastroNeto2009,Novoselov2005,Avouris2010,Balandin2008}. Its zero (direct) band gap restricts the usage to field-effect transistors (FETs) that require electronic switching properties \cite{Xia2010,Schwierz2010,Han2007}.
The direct band gap of transition metal dichalcogenides (TMDs), which ranges from 1 to 2 eV, has made them popular among researchers studying materials that can replace graphene \cite{Huo2017,Podzorov2004,Li2015}. TMDs consist of atomically thin X--M--X layers which contain a transition metal (M) layer that is encased between two chalcogen atom (X = S, Se, or Te) layers  \cite{Wang2012,Chhowalla2013}. Each monolayer adopts a hexagonal geometry characterized by strong intraplanar covalent bonds \cite{Wang2012,Manzeli2017} while adjacent layers are held together by weak van der Waals interactions \cite{Wang2012,Chhowalla2013,Manzeli2017,Coleman2011}.
The layered structure of TMDs enables their materials system to display strong anisotropy \cite{Zhao2013,Li2013}, which allows exceptional mechanical flexibility \cite{Bertolazzi2011,Cooper2013} and permits the extraction of single-layer material through exfoliation methods \cite{Novoselov2005,Coleman2011}. The materials exhibit different crystallization patterns which depend on their atomic coordination and crystal stacking, resulting in three distinct structural phases that include the $2$H, $1$T, and $1$T' phases \cite{Voiry2015, Eda2012, Lin2014}. Tungsten diselenide (WSe$_2$) belongs to the TMDs family because it exhibits strong spin–orbit coupling \cite{Zhu2011} and produces carriers that have both spin and valley polarization characteristics \cite{Xu2014,Wang2015}. The chemical and mechanical stability of this material enables its use in heterostructures and flexible devices \cite{Bertolazzi2011,Cooper2013}, while its layered structure enables monolayer material extraction through exfoliation methods that create substantial material anisotropy \cite{Novoselov2005,Coleman2011}. The electronic properties of two-dimensional systems change according to structural and strain effects which make WSe2 an appropriate material for research purposes.

{Strain engineering has proven to be a successful method of tailoring electronic and optical properties of WSe$_2$, because even slight lattice deformations can dramatically affect charge carrier transport~\cite{Aslan2020,Conley2013Biaxial,Deng2020}. This is achieved through either tensile or compressive strain, which modifies interatomic distance and affects electronic band structure accordingly. Uniaxial strain can be applied by stretching or compressing a material along a certain crystallographic direction with the help of elastic substrates or mechanical devices~\cite{Cui2014,Lui2014,Han2015}.
	Other popular methods involve deposition of WSe$_2$ monolayer on flexible substrates, such as PDMS~\cite{Kim2010,Zhou2015}. By bending or stretching the substrate, one can introduce controllable and almost uniform strain into a monolayer. With this approach, it becomes possible to reversibly modulate the band gap, phonon spectra, and carrier mobility, as was shown in experiments by Yang~\textit{et al.}~\cite{Yang2021} and Peng~\textit{et al.}~\cite{Peng2020}. Furthermore, wrinkling of a pre-strained substrate results in the appearance of localized strain fields, which allow trapping charge carriers and modifying the band gap locally~\cite{He2015,Kumar2015,Branny2017,Palacios2017}.
	In addition to mechanical strain, biaxial strain can be introduced through hydrostatic pressure or by using thermal expansion mismatch between monolayer and substrate~\cite{Zhang2020ExtremePressure,Conley2013Biaxial}.}

The study  of TMDs, especially monolayer and multilayer WSe$_2$, has attracted considerable attention in recent years~\cite{Chhowalla2013,Wang2015,MaterialSciRep2018,OptoElec2024}.  Several theoretical and experimental investigations have demonstrated that strong spin-orbit coupling and valley symmetry play {an} essential role in modulating electronic transport, with energy- and valley-dependent tunneling phenomena being observed~\cite {Xiao2012,Kormanyos2015kp}. Furthermore,  carrier transport in field-effect transistors and heterostructures based on WSe$_2$ exhibits ambipolar behavior~\cite{Seo2019Ambipolar,PolarityControl2014}, high on/off ratios~\cite{HighIonIoff2017,HighOnOffWSe2_AdvMat}, and efficient conductance modulation through electrostatic doping~\cite{ElectricDoubleLayer2016,Ghosh2025HighPerf}. The influence of electric, magnetic, and optical fields has also been extensively investigated, highlighting  the control of  spin and valley polarizations, paving the way for valleytronic devices~\cite{Mak2012Control,Sanchez2016ValleyPolarization,Schaibley2016Valleytronics}. Although this subject has been extensively studied, we believe that the combination of strain and scalar potential in WSe$_2$ could reveal interesting aspects.

We focus on the analysis of electronic transport in a WSe$_2$ monolayer under uniaxial strain and a scalar potential. We solve the Dirac equation to derive the eigenvalues and eigenvectors for each region.  Using the continuity conditions at interfaces, we explicitly determine the transmission, the conductance, and the spin and valley polarizations as functions of system parameters. As a result, we show that the behavior of WSe$_2$  is affected by strain, barrier parameters, incident energy, and incidence angle. Indeed,  it is found that at normal incidence, transmission can reach nearly unity, showing Klein tunneling persists despite the spin–orbit–induced energy gap, while increasing the incidence angle reduces transmission, vanishing beyond a critical value determined by strain and incident energy. Further, variations in the barrier parameters lead to oscillations in transmission and conductance, characteristic of Fabry–Pérot–type quantum interference, accompanied by spin- and valley-polarized transport, which can be tuned through the selective separation of transport channels.

The remainder of our paper is organized as follows. In Sec.~\ref{sec1}, we introduce the theoretical framework by presenting the Hamiltonian of the system and solving the Dirac equation to derive the corresponding energy spectrum. In  Sec.~\ref{sec2}, we impose the continuity of the spinor wave functions at the interfaces $x=0$ and $x=D$  to determine the transmission coefficient. Next,  we analyze the transmission probability behavior numerically under various conditions. {In Sec.~\ref{cond}}, we compute the corresponding conductance  using the Büttiker formalism. This allows us to analyze  spin and valley polarizations in Sec.~\ref{pol}. We discuss the underlying physical mechanisms { {in Sec.~\ref{phy}}. Finally, we  conclude our work by summarizing the main results.

% The present paper is organized as follows. In Sec.~\ref{sec1}, we establish the theoretical model by introducing the Hamiltonian describing the system analyzed and solving the Dirac equation to obtain the energy spectrum. The matching of the spinors at the interfaces $x=0$ and $x=D$ allows us to calculate the transmission, and  numerically analyzed under various conditions in Sect.~\ref{sec2}. In Sec.~\ref{con}, the conductance is evaluated using the Büttiker relation. Sec.~\ref{pol} addresses polarization, while  Sec.~\ref{phy} provides a detailed physical interpretation. Our conclusions are presented in  Sec.~\ref{con}.

\section{Theoretical model}\label{sec1}
%We consider a WSe$_2$-based system composed of three regions labeled \(j = 1, 2, 3\).
% Regions 1 and 3 correspond to normal (unstrained)  monolayer WSe$_2$, while region $2$
% consists of  WSe$_2$ subjected to uniaxial strain and an external scalar electrostatic potential applied along the transport direction, as schematically illustrated in Fig.~\ref{str}.
We study the transport properties of a monolayer WSe$_2$ consisting of three consecutive regions, denoted by \(j = 1, 2, 3\). Regions 1 and 3 correspond to pristine (unstrained) monolayer WSe$_2$ and serve as the source and drain leads, respectively.  
%Region 2, located between them, is formed by monolayer WSe$_2$ subjected to both uniaxial strain and an externally applied scalar electrostatic potential along the transport direction. 
{Region 2, located between them, is formed by a monolayer of WSe$_2$ subjected to both a scalar electrostatic potential generated by two electrodes connected to a static generator \cite{staticbarr} and a uniaxial strain created by depositing the sheet on a flexible substrate \cite{strain, strain2}.} This central strained region acts as a tunable barrier that modifies the local electronic structure and carrier propagation. The applied uniaxial strain changes the lattice geometry and consequently alters the band dispersion, while the scalar potential shifts the energy landscape within the barrier. The combined action of these two mechanisms produces a controllable scattering region sandwiched between the unstrained leads.
{The barrier height is controlled by an external electrostatic potential, while the uniaxial strain is introduced mechanically through substrate engineering. Since they affect different physical properties, these two parameters can be tuned independently and are therefore treated as independent in our model.} A schematic representation of the system is shown in Fig.~\ref{str}.
	\begin{figure}[H]
		\centering
		\includegraphics[scale=0.3]{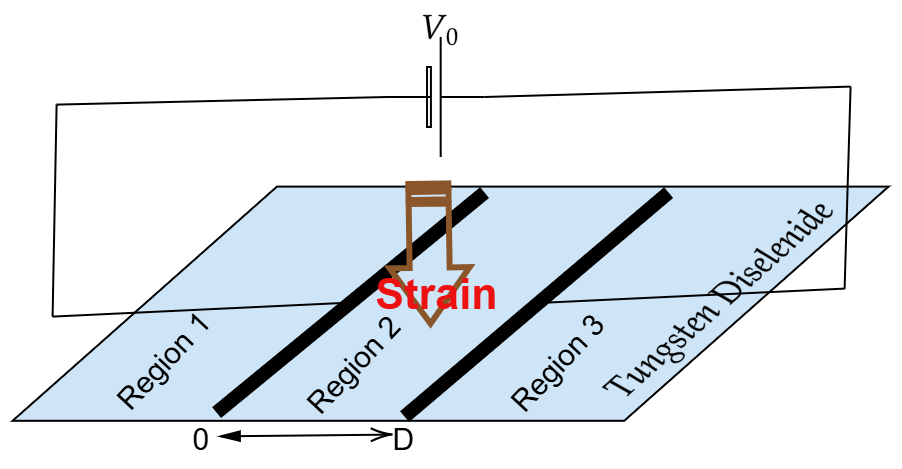}
        \caption{
        Schematic representation of a WSe$_2$ sheet subjected to a uniaxial strain along with an external scalar electrostatic potential of height $V_0$ and width $D$ in the central region.}
        %, which separates the system into three distinct regions.}
        \label{str}
	\end{figure}

%\section{Low-energy Hamiltonian}

The low-energy effective Hamiltonian of a strained monolayer WSe$_2$ near the two inequivalent valleys \(K\) and \(K'\) is given by
\begin{widetext}
    \begin{align}\label{Ham}
		H =  v_F \left( \tau \sigma_x p_x + \sigma_yp_y\right) + \frac{\Delta}{2}\sigma_z -\lambda_c \tau s_z\frac{(\sigma_0+\sigma_z)}{2}- \lambda_v\tau s_z\frac{(\sigma_0-\sigma_z)}{2}+V(x)\sigma_0+H_{strain}
	\end{align}
\end{widetext}
where  the Fermi velocity is $v_F =0.5\times10^{6}$ \text{m/s}, the intrinsic band gap is
\(\Delta~=~1.7~\text{eV}\). The spin--orbit coupling strengths are
  $\lambda_v=0.1125$ \text{eV} for valence band and $\lambda_c=0.075$ \text{eV} for conduction band  \cite{WSe2}.
\(s_z=\pm1\) labels the electron spin, \(\sigma_i\) are Pauli matrices acting in the pseudospin space. The scalar electrostatic potential forms a rectangular barrier along the $x$-direction, such as
\begin{align}
V(x)=V_0\left[\Theta(x)-\Theta(x-D)\right]
\end{align} where $V_0$ denotes the barrier height and $\Theta(x)$ is the Heaviside step function. In a monolayer WSe$_2$, uniaxial strain alters the low-energy electronic properties through an effective valley-dependent gauge field, which enters the Hamiltonian as

%\subsection{Strain-Induced Terms}
%Mechanical strain in a monolayer MoS$_2$ modifies the electronic spectrum through the emergence of an effective gauge field that couples to the Dirac fermions in a valley-dependent manner.
%Uniaxial strain modifies the Hamiltonian through a valley-dependent gauge field,
\begin{align}
	H_{\text{strain}} &= \hbar v_F \left( \tau \sigma_x A_x + \sigma_y A_y \right)
\end{align}
and the deformation-induced vector potential can be written
\begin{align}
	A_x = \beta \left( u_{xx} - u_{yy} \right), 
	\quad
	A_y = -2 \beta u_{xy}
\end{align}
where \(u_{ij}\) refers to the strain tensor components and \(\beta ~\approx~2.4\) is the Gr\"uneisen parameter specific to the material \cite{beta}. 
In this work, we consider uniaxial strain applied along the transport direction $x$, such that $u_{xy}=0$ and only the $A_x$ component remains nonzero.  For a uniform deformation, the strain tensor components are given by
\begin{align}
	u_{xx} = \varepsilon, 
	\quad
	u_{yy} = -\nu \varepsilon
\end{align}
assuming \(\varepsilon\) is the strain amplitude and \(\nu = 0.25\)  is the Poisson ratio \cite{nu}, the gauge field simplifies to a constant
\begin{align}
	A_x = \beta \varepsilon (1+\nu).
\end{align}
In matrix representation, the Hamiltonian (\ref{Ham}) takes the form
\begin{align}
    H = \begin{pmatrix} V_c  &  v_{F}\left[\tau(p_{x}+\hbar A_{x}) - ip_{y}\right] \\  v_{F}\left[\tau(p_{x}+\hbar A_{x}) + ip_{y}\right] & V_v \end{pmatrix}
\end{align}
where the diagonal elements \(V_c\) and \(V_v\) represent the energy levels of the conduction and valence bands, respectively, including contributions from the intrinsic gap and spin-orbit coupling. We define these quantities as
\begin{align}
    &V_c=V_0+	\frac{\Delta}{2} - \lambda_c \tau s_z \\
    &V_v=V_0-\frac{\Delta}{2} +\lambda_v \tau s_z.
\end{align}

The dynamics of the charge carriers are governed by the stationary Dirac equation
    $H\,\psi_j(x,y) = E\,\psi_j(x,y),
$
which determines the allowed energy eigenvalues \(E\) and their corresponding spinor solutions \(\psi_j\). Owing to translational invariance  along the \(y\) direction, we can separate the variables and write the wave function in the form
$\psi_j(x,y) = e^{ik_y y}\phi_j(x).$
In  region 2, the longitudinal wave vector \(q'_x\)  satisfies the dispersion relation 
\begin{align}
	(E - V_c)(E-V_v) =
	\hbar^2 v_F^2 \left[ (q'_x +A_x)^2 + k_y^2 \right].
\end{align}
The evolution of the spinor components is then governed by the associated eigenvalue equations. Here, the two‐component spinor is defined as $\phi (x)= 
	\begin{pmatrix}
		\psi_{A}(x),
		\psi_B(x)
	\end{pmatrix}
	^T$, with $T$ stands for transpose.  Substituting this form into the stationary Dirac equation yields the coupled first-order equations
\begin{align}\label{eq1}
  &  \left(E - V_c\right)\psi_{A} = \hbar v_{F}\left[\tau(-i\partial_{x}+A_x)- ik_{y}\right]\psi_{B}\\ 
 & \left(E - V_v\right)\psi_{B} = \hbar v_{F}\left[\tau(-i\partial_{x}+A_x)+ ik_{y}\right]\psi_{A}\label{eq2}.
\end{align} 
Decoupling (\ref{eq1}) and (\ref{eq2}) gives
\begin{align}
\left[\partial_{x}^{2} +2i A_x\,\partial_{x}+\frac{(E - V_c)(E - V_v)}{(\hbar v_{F})^{2}}-k_y^2\right]\psi_{A} = 0
\end{align}
%where$q_x^2=\frac{(E - V_c)(E - V_v)}{(\hbar v_{F})^{2}}-k_y^2$\\
which describes the effective one‑dimensional propagation in the \(x\) direction. The analytical solution of this decoupled equation yields the following spinor
\begin{align}
   \psi_{2} = e^{-i\tau A_{x}x}\left[a_{1}\begin{pmatrix}
      1\\\beta'
 \end{pmatrix}e^{i q'_x x} + a_{2}\begin{pmatrix}
      1\\-\beta'^*
  \end{pmatrix}e^{-i q'_x x}\right]e^{ik_yy}
\end{align}
where the wave vector ${q_x'}^{2}$ and the complex number $\beta'$ are
\begin{align}
    {q_x'}^{2} &= \frac{(E - V_c)(E - V_v)}{(\hbar v_{F})^{2}} - k_y^{2} - A_x^{2} \\
    \beta'&= \dfrac{\hbar v_{F}\left[\tau\left(q_x' +  A_x\right)+ i k_{y}\right]}{E - V_v}
    = e^{i\theta}
\end{align} with the angle $\theta=\text{tan}^{-1}\!\left(\dfrac{k_y}{\tau\left({q_x'} +A_x\right)}\right)$.
%with $q'^2=q_x^2-A_x^2$ and $\beta=\frac{\hbar v_{F}\left[\tau q'_x+ i k_{y}\right]}{E - V_v}=se^{i\theta}$.\\
In region 1 ($x<0$), we have prestine WSe$_2$ and then the spinor is
\begin{align}
    \psi_1(x,y) = \left[\begin{pmatrix}
        1\\\gamma
    \end{pmatrix}e^{i k_x x} + r\begin{pmatrix}
        1\\-\gamma^*
    \end{pmatrix}e^{-i k_x x}\right]e^{ik_yy}
\end{align} 
where  $r$ is  the reflection  coefficient, the wave vector $k^2_x$ and the complex number $\gamma$ are
\begin{align}
&k^2_x=\frac{(E-\Delta-s_z \tau\lambda_c)(E+\Delta-s_z \tau\lambda_v)}{\left(\hbar v_{F}\right)^2}-k_y^2\label{kxx}\\
    &\gamma=\frac{\hbar v_{F}\left[\tau k_x+ i k_{y}\right]}{E -\frac{\Delta}{2} +\lambda_v \tau s_z}=e^{i\phi}
\end{align}
$\phi=\text{tan}^{-1}\!\left(\dfrac{k_y}{\tau{k_x}}\right)$ is the incident angle.
In region 3 ($x>D$), we have just the transmit wave described by the spinor
\begin{align}
    \psi_3(x,y)=t\begin{pmatrix}
        1\\ \gamma
    \end{pmatrix}e^{ik_xx}e^{ik_yy}
\end{align}
with  the  transmission coefficient $t$. 
{The energy spectrum obtained from these solutions will be used to analyze the transport properties of the system. Specifically, the eigenvalues and corresponding wave functions in each region serve as the building blocks for constructing the transfer matrix, which connects the wave amplitudes across the barrier interfaces. The transmission and reflection coefficients are then derived from the boundary conditions at the interfaces between the strained and unstrained regions. These quantities are subsequently used to calculate the conductance, as well as the spin and valley polarizations, as functions of the relevant physical parameters, including strain, barrier height, barrier width, and incident energy.}

\section{Transmission}\label{sec2}
To determine the reflection and transmission coefficients, we apply the boundary conditions at the interfaces between the different regions. Explicitly, we impose  $\psi_{1}(0,y) = \psi_{2}(0,y)$ at the first interface and $ \psi_{2}(D,y) = \psi_{3}(D,y)$ at the second one. Since each wavefunction is a two-component spinor, these continuity conditions lead to a system of four coupled equations, which can be solved to obtain the reflection and transmission coefficients.
\begin{align}
    &1+r=a_1+a_2\\
    &\gamma-r\,\gamma^*=\beta' a_1- \beta'^* a_2\\
    &e^{-i \tau A_x D}(a_1e^{iq'_xD}+a_2e^{-i\tau q_xD})=t\,e^{ik_xD}\\
    &e^{-i \tau A_x D}(a_1\beta'\,e^{iq'_xD}-a_2\beta'^* e^{-iq'_xD})=t\gamma\,e^{ik_xD}
\end{align}
Solving this system allows us to obtain the reflection and transmission amplitudes
\begin{align}
   &r = i\,e^{i\theta}\frac{\big(e^{2 i q'_{x}D}-1\big)\sin\!\left(\tfrac{\theta-\phi}{2}\right)\, \cos\left(\tfrac{\theta+\phi}{2}\right)} {e^{2 i q'_xD}\sin^{2}\left(\tfrac{\theta-\phi}{2}\right) - \cos^{2}\left(\tfrac{\theta+\phi}{2}\right)}\\
    &t = \frac{e^{i\phi}\,e^{-iD(\tau A_{x}+k_{x}-q'_{x})}\cos\theta}{- e^{2iDq'_{x}}\sin^{2}\left(\tfrac{\theta-\phi}{2}\right) + e^{i\phi}\cos\left(\tfrac{\theta+\phi}{2}\right)\cos\left(\tfrac{\theta-\phi}{2}\right)}
\end{align} 
By using the continuity equation to compute the current density, we ensure conservation of probability flux across all regions, which provides the expressions for the reflection and transmission probabilities
\begin{align}
T &= \frac{J_{\text{tran}}}{J_{\text{inc}}} = |t|^2, \qquad
R &= \frac{J_{\text{ref}}}{J_{\text{inc}}} = |r|^2.
\end{align}
Algebraic simplification then yields
\begin{align}
T=\frac{\cos^{2}\theta\,\cos^{2}\phi}
{\cos^{2}\theta\,\cos^{2}\phi\,\cos^{2}(q'_x D)
+ \sin^{2}(q'_x D)\left(1-\sin\theta\,\sin\phi\right)^{2}}.
\end{align}
This expression highlights the crucial role of quantum interference and the angle of incidence in transport, and shows that transmission can become perfect for certain values of $\theta$ and $D$, a phenomenon characteristic of relativistic tunneling (the Klein effect). To illustrate and quantitatively analyze these analytical results, we now turn to a numerical study of the transmission probability. This will allow us to explore the influence of the system’s physical parameters—such as strain, the scalar potential, the angle of incidence, the barrier width, and the incident energy—on transport behavior.

\begin{figure} [H]
	\centering
\subfloat[$D=1$ nm]{\includegraphics[scale=0.27]{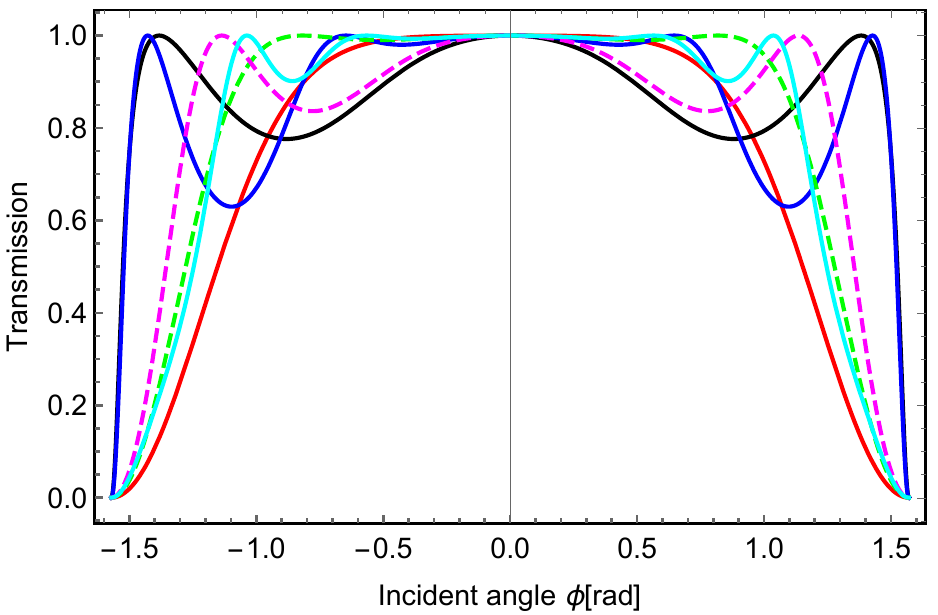}}\label{phia}\subfloat[$D=2$ nm]{\includegraphics[scale=0.27]{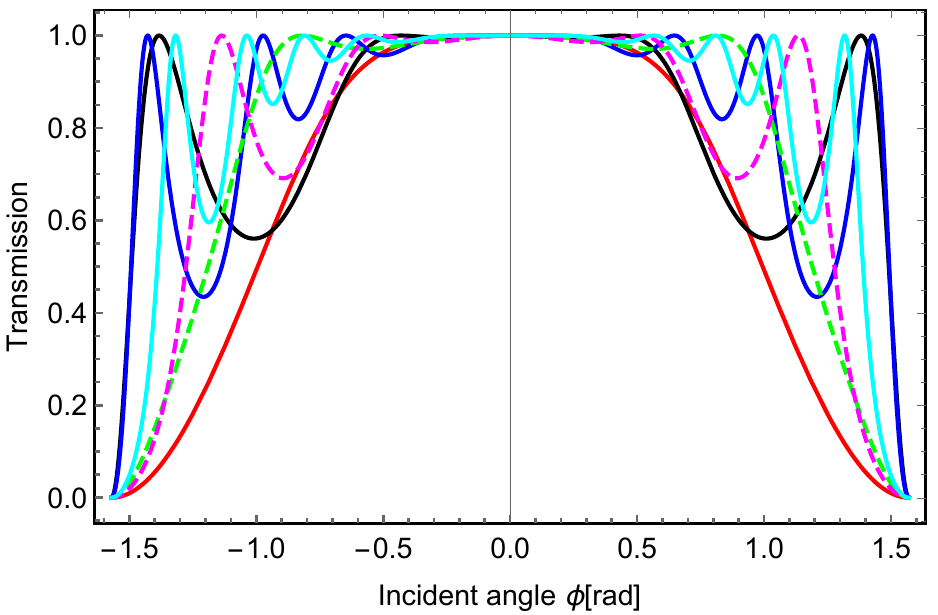}}\label{phib}
\subfloat[$D=5$ nm]{\includegraphics[scale=0.27]{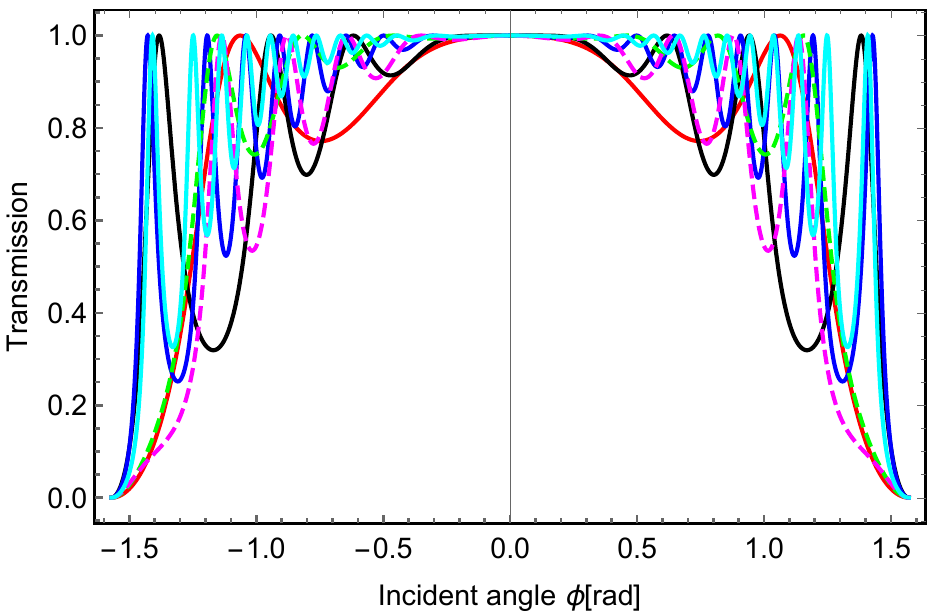}}\label{phic}
		\caption{Transmission probability in the $K$ valley as a function of the incident angle $\phi$, for spin-up (solid line) and spin-down (dashed line). Three incident energies are considered: 0.9 eV (red, green), 1.1 eV (black, magenta), and 1.3 eV (blue, cyan), with $\varepsilon=1$ and $V_0=3$ eV.}\label{phi}
%Transmission of $k$ valley as a function of incident angle $\phi$, for $\varepsilon=1$ and $V_0=2$ eV. $E=0.6$ eV red (green dashed)line, $E=0.8$ eV black (magenta dashed) line and $E=1$ eV blue (cyan dashed) line. Spin up solide line and spin down dashed line}\label{phi}
        \end{figure}

Figure~\ref{phi} displays the transmission probability in the $K$ valley as a function of the incident angle $\phi$ for different barrier widths $D$ and incident energies $E$, under finite uniaxial strain. A prominent feature is the near-unity transmission at normal incidence $(\phi=0)$ for all considered parameters, which constitutes a clear signature of Klein tunneling in monolayer WSe$_2$, similar to the behavior previously reported in graphene-based Dirac systems \cite{Katsnelson2006,CastroNeto2009}. Despite the presence of a finite band gap induced by intrinsic spin--orbit coupling, the chiral nature of Dirac fermions allows perfect transmission when pseudospin matching across the barrier is preserved, demonstrating that Klein's tunnel effect remains robust even in Dirac band gap materials \cite{Gu2011,Wu2018}.
As the incident angle increases, the transmission progressively decreases and eventually vanishes beyond a critical angle $\phi_{\mathrm{max}}$, which depends on the incident energy and the strain strength. This angular cutoff arises from the conservation of the transverse momentum $k_y$, which restricts the existence of propagating modes inside the barrier and leads to evanescent states at larger angles, in agreement with earlier theoretical studies \cite{Pereira2006}. Increasing the barrier width strongly enhances oscillatory patterns in the transmission, reflecting Fabry--Pérot–like interference due to multiple reflections between the barrier interfaces \cite{Zarenia2013}. These oscillations become denser at higher energies, consistent with the increase of the longitudinal wave vector inside the barrier.
The small but finite difference between spin-up and spin-down transmission originates from spin--orbit coupling, which slightly shifts the effective band edges and modifies the phase accumulation across the barrier, leading to spin-dependent transport characteristics as predicted for monolayer {TMDs} \cite{Xiao2012,Liu2015,Wu2018}.  These results highlight the combined influence of uniaxial strain and intrinsic spin--orbit coupling on spin-resolved ballistic transport in WSe$_2$ \cite{Rostami2015,Zarenia2013}.
%This indicates that even at the transmission level, strain and spin--orbit coupling jointly influence spin-resolved transport.

\begin{figure}[ht]
	\centering
	\subfloat[$\phi=\pi/6$]{\includegraphics[scale=0.27]{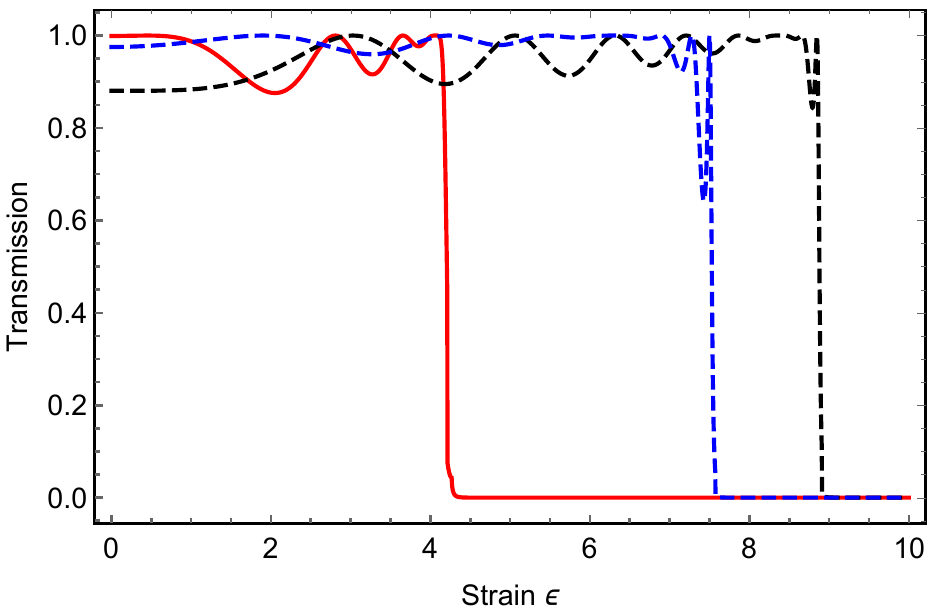}}\label{phia}\subfloat[ $\phi=\pi/4$]{\includegraphics[scale=0.27]{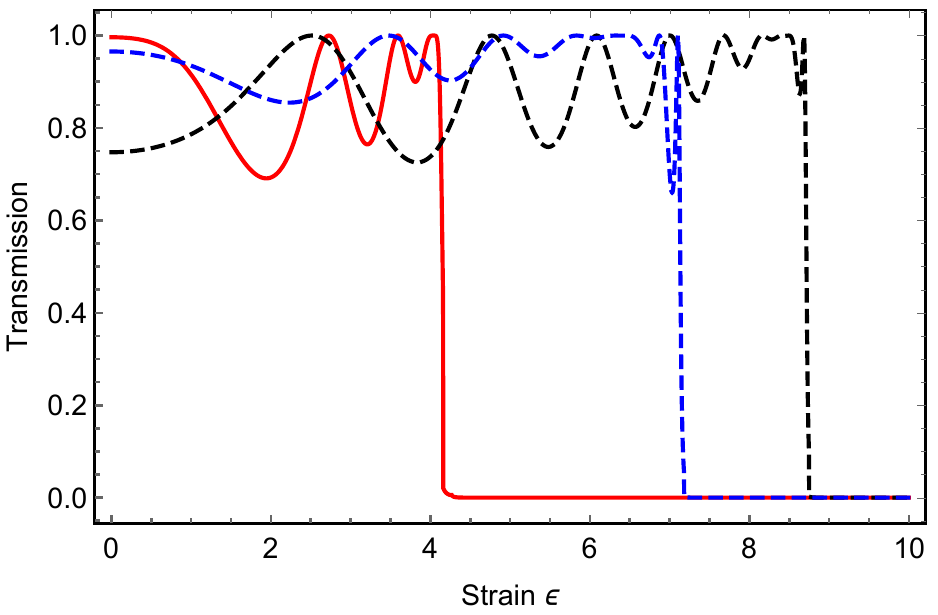}}\label{phib}
    \subfloat[$\phi=\pi/3$]{\includegraphics[scale=0.27]{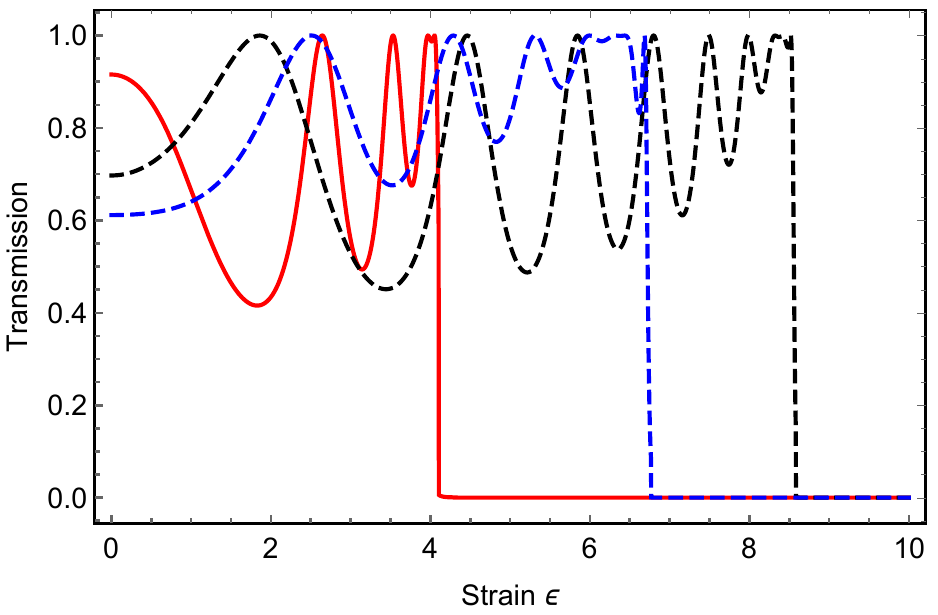}}\label{phic}
	\caption{Spin-up transmission probability in the $K$ valley as a function of strain $\varepsilon$ for three different incident energy values;  $E=1.1$ eV (black dashed line), $E=1.3$ eV (blue dashed line) and $E=1.5$ eV (red line), with $D=1$ nm and  $V_0=3$ eV.}\label{epsilon}
 % Transmission (spin up) as a function of strain $\varepsilon$, for $D=5$ nm and  $V_0=2$ eV. $E=0.6$ eV (red line), $E=0.8$ eV (black dashed line) and $E=1$ eV (blue dashed line)}\label{epsilon}
\end{figure}

%Figure~\ref{epsilon}  shows the spin-up transmission probability, corresponding to the \(K\) valley, as a function of uniaxial strain $\varepsilon$ for fixed barrier width and several incident angles. 
Figure~\ref{epsilon} shows the spin-up transmission probability, corresponding to the \(K\) valley, as a function of uniaxial strain $\varepsilon$ for fixed barrier width and several incident angles. 
{In the absence of strain ($\varepsilon=0$), the resonance pattern is fully determined by the electrostatic barrier geometry and the intrinsic band structure of monolayer WSe$_2$, where interference effects arise solely from energy- and width-dependent phase accumulation inside the barrier. When uniaxial strain is applied, a valley-dependent gauge field is generated, which effectively shifts the longitudinal momentum inside the barrier region. Consequently, the phase accumulated across the barrier becomes strain-dependent, leading to a modification of the resonance condition $q_x' D = n\pi$. This gives rise to  well-defined oscillations in the transmission as strain varies, as observed in \cite{Rostami2015,Zarenia2013}.} 
For small incident angles, transmission remains relatively robust against strain variations, reflecting a low sensitivity of the coupling between the transverse momentum and the strain-induced effects. In contrast, for larger angles, transmission oscillations become more pronounced, revealing a stronger interaction between the transverse component of the wave vector and the strain-induced shift of the longitudinal wave vector. This angular dependence indicates that strain acts as an efficient external control knob to dynamically modulate electronic transmission through the barrier \cite{Roldan2013,Guinea2010}. Moreover, at sufficiently high strain values, the transmission can approach unity again, demonstrating that strain can restore resonant tunneling even when it is suppressed at lower strain. The ability to restore resonant transmission through strain engineering is a major advantage for the development of nanoelectronic devices based on WSe$_2$ or other $2$D materials, where valley-spin coupling and mechanical sensitivity can be exploited \cite{Zarenia2013,Manzeli2017}.

\begin{figure}[ht]
	\centering
	\subfloat[$\phi=\pi/6$]{\includegraphics[scale=0.27]{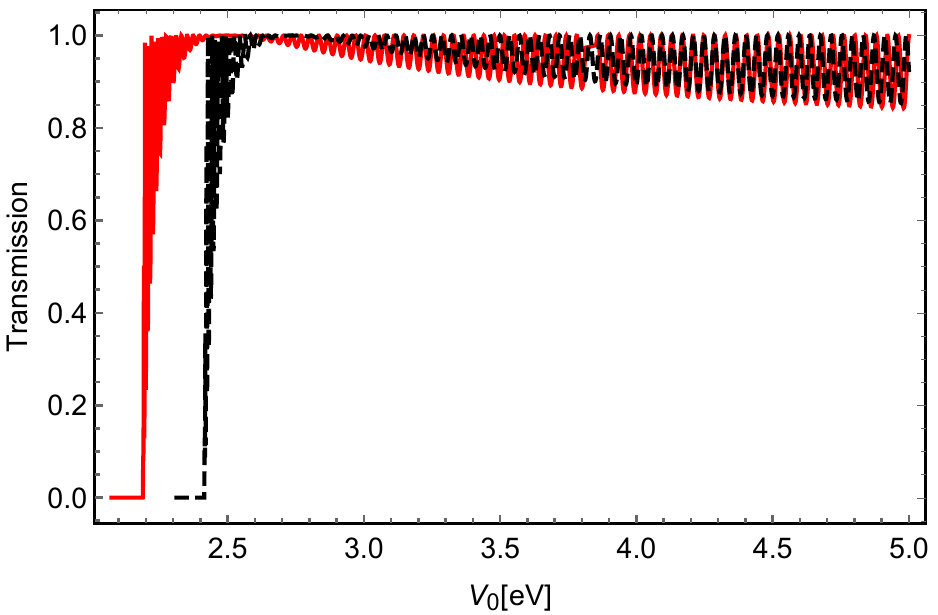}}\label{TVa}\subfloat[ $\phi=\pi/4$]{\includegraphics[scale=0.27]{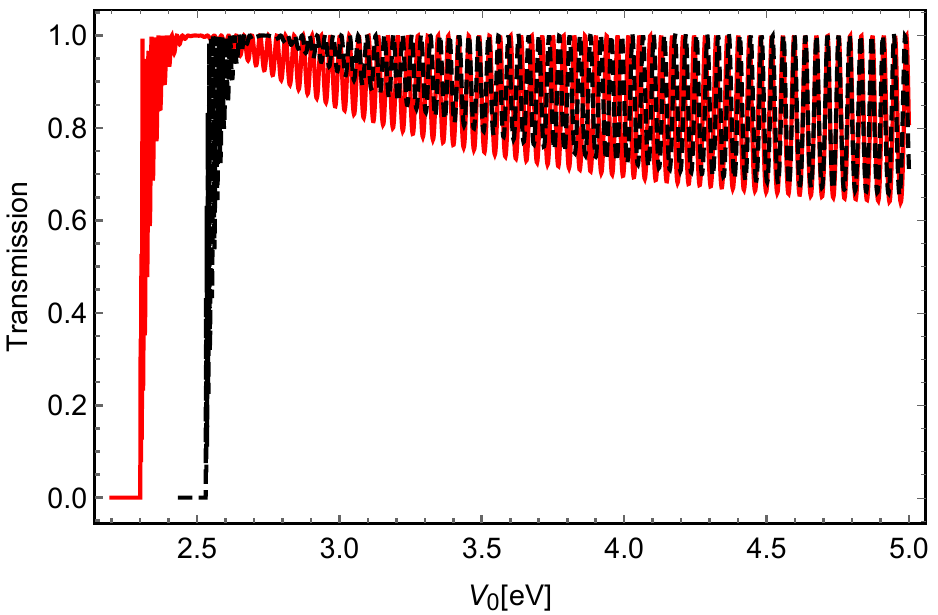}}\label{TVb}
	\subfloat[$\phi=\pi/3$]{\includegraphics[scale=0.27]{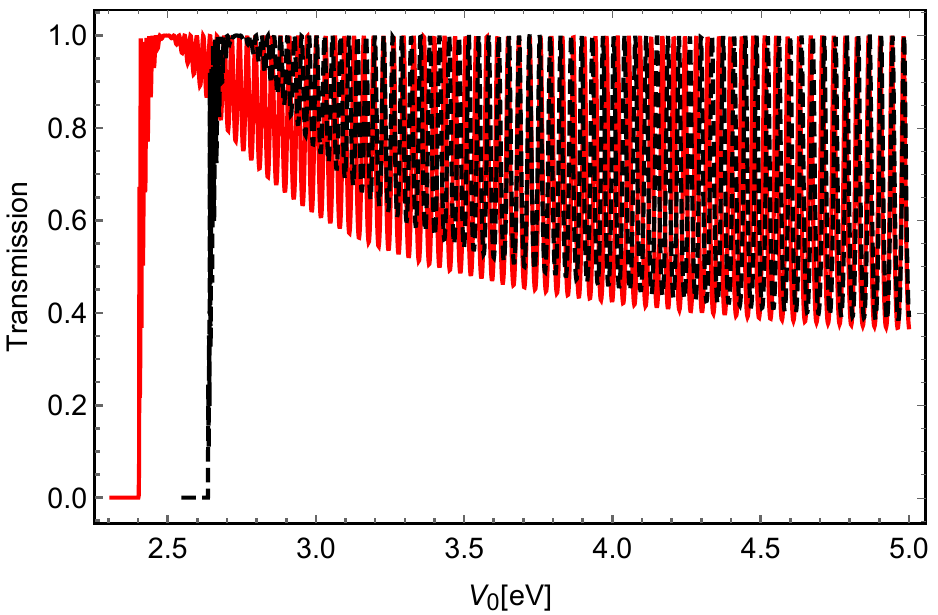}}\label{TVc}
	%\caption{Transmission as a function of barrier height $V_0$ for $D=5$ nm and $V_0=2$ eV. Spin down (black line) and spin up (red line)}
    \caption{Spin-up transmission (black line) and spin-down transmission (red line)  in the $K$ valley as a function of the barrier height $V_{0}$ for $D =5$ nm, $\varepsilon=1$ and $E=1.3$ eV.}\label{TV0}
\end{figure}

Figure~\ref{TV0} shows the transmission probability as a function of the barrier height $V_0$ for various incident angles. A sharp threshold behavior is found: for barrier heights less than a critical value the transmission exhibits strong suppression because there are no propagating states inside the barrier. Beyond this limit, oscillatory transmission is observed, which signals the appearance of quasibound states and resonant tunneling phenomenon Physically, the oscillations are due to constructive and destructive interference of the electron wave function inside the barrier  reminiscent of Fabry-Pérot resonances found in other two-dimensional materials \cite{Rostami2015,Zarenia2013,Kormanyos2015kp}.
The threshold value grows with the increase of incident angle, indicating that less longitudinal kinetic energy is available for tunneling. In addition, the amplitude of oscillations increases with $V_0$, due to stronger confinement which promotes interference. The splitting between spin-up and spin-down curves is a direct evidence that spin--orbit coupling leads to propagation thresholds depending on the electron's spin. This splitting reflects the fact that electrons with different spin preference feel dissimilar effective barrier potentials, which are furthermore tunable by strain and electro static gating. These features make the system a promising platform for the realization of strain-engineered spintronic and valleytronic devices, such as spin filters, valley polarizers, and electrically tunable quantum interference transistors \cite{Wang2015,Yang2025,NanoEnergy2020,Zhang2026,PhysicaE2024}.

\section{Conductance}\label{cond}
% To connect microscopic quantities to macroscopic ones, the conductance is calculated. By definition, at zero temperature the conductance is given by the integral of the total transmission over $k_y$ \cite{Landau,Mandal}; at the same time, it corresponds to the average fermion flux across half of the Fermi surface \cite{butker, conduct1}. This approach links quantum transport to classical concepts such as current density and Ohm's law, showing that macroscopic linear-response behavior emerges from the sum of microscopic transmission channels. Furthermore, the conductance can be related to other physical quantities, notably the density of states and the group velocity of charge carriers, which appear directly when calculating the flux through a scattering region. This is

To bridge microscopic transport processes with macroscopic observables, we evaluate the conductance of the system. At zero temperature, the conductance is defined as the integral of the total transmission probability over the transverse wave vector $k_y$ \cite{Landau,Mandal}. Equivalently, it can be interpreted as the average fermionic flux across one half of the Fermi surface \cite{butker,conduct1}. This formulation provides a direct connection between quantum transport and classical concepts such as current density and Ohm’s law, illustrating how macroscopic linear-response behavior emerges from the collective contribution of microscopic transmission channels.
Moreover, the conductance can be expressed in terms of fundamental quantities such as the density of states and the group velocity, which naturally arise when computing the particle flux through the scattering region. It is given by
\begin{align}
	G_\tau = \frac{G_0}{2\pi} \int_{-k_y^{\text{max}}}^{k_y^{\text{max}}} (T_{\uparrow\tau}(E, k_y)+T_{\downarrow\tau}(E, k_y))\, dk_y	
\end{align}
% where \(G_0=\frac{2\pi e^2}{\hbar}\) is the unit of conductance, and \(k_y^{\text{max}}\) represents the maximum wave vector component along the \(y\)-direction. The connection between the transverse wave vector \(k_y\) and the incident angle \(\phi\)  allows the conductance \(G\) to be expressed as a sum of angular contributions, thereby making the correspondence with semi-classical trajectories explicit. This formulation also emphasizes the role of angle-dependent scattering, reminiscent of classical transport phenomena such as anisotropic conductivity and the Boltzmann transport equation:
with \(G_0 = \frac{2\pi e^2}{\hbar}\) is the conductance quantum,  \(k_y^{\text{max}}\) is  the maximum transverse wave vector \(k_y\). Using the relation between  \(k_y\) and the incident angle \(\phi\),  the conductance \(G\) can be reformulated as an intact over angular contributions. This representation makes the link with semiclassical carrier trajectories further obvious and highlights the significance of angle dependent scattering, in close analogy with classical transport mechanisms such as anisotropic conductivity described within the Boltzmann transport framework. This is
\begin{equation}
	G_\tau = \frac{G_0}{2\pi} \int_{-\phi^{\text{max}}}^{\phi^{\text{max}}} (T_{\uparrow\tau}(E, \phi)+T_{\downarrow\tau}(E, \phi)) \cos\phi \, d\phi_0	
\end{equation}
and the maximum value of \(\phi_0^{\text{max}}\) can be determined through the equation \(k_y^{\text{max}} = k \sin\phi_0^{\text{max}}\), which uses \(k\) from \eqref{kxx}. 
% and \(\phi_0^{\text{max}}\) can be obtained from the relation \(k_y^{\text{max}} = k \sin(\phi_0^{\text{max}})\), with \(k\) given by \eqref{kxx}.  This establishes a link between the energy $E$ and the maximum angle of incidence $\phi_0^{\text{max}}$, determining the modes that effectively contribute to the conductance. The angular dependence affects the transmission probabilities, while the spin- and valley-resolved channels reveal microscopic effects in macroscopic measurements.
Since the transmission probabilities depend explicitly on the incident angle, the resulting conductance captures the intrinsic scattering anisotropy of the system. Furthermore, the spin- and valley-resolved transmission channels encode microscopic quantum effects that are directly reflected in measurable macroscopic transport properties.

%This relation provides a connection between the energy \(E\) and the maximum angle \(\phi_0^{\text{max}}\) for the allowed transmission channels.
\begin{figure}[H]
	\centering
	\subfloat[$E=1.1$ eV]{\includegraphics[scale=0.27]{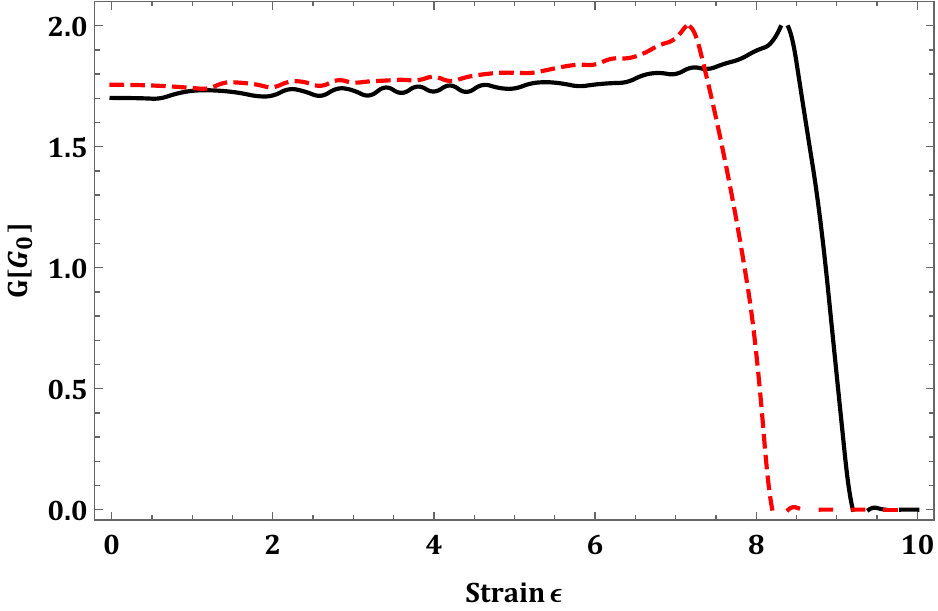}}\label{phia}\subfloat[ $E=1.3$ eV]{\includegraphics[scale=0.27]{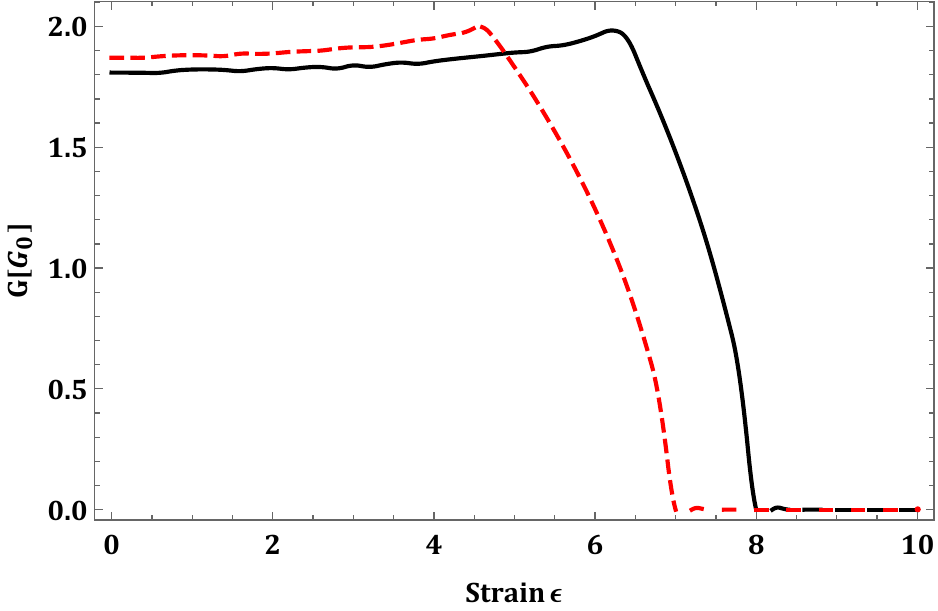}}\label{phib}
	\subfloat[$E=1.5$ eV]{\includegraphics[scale=0.27]{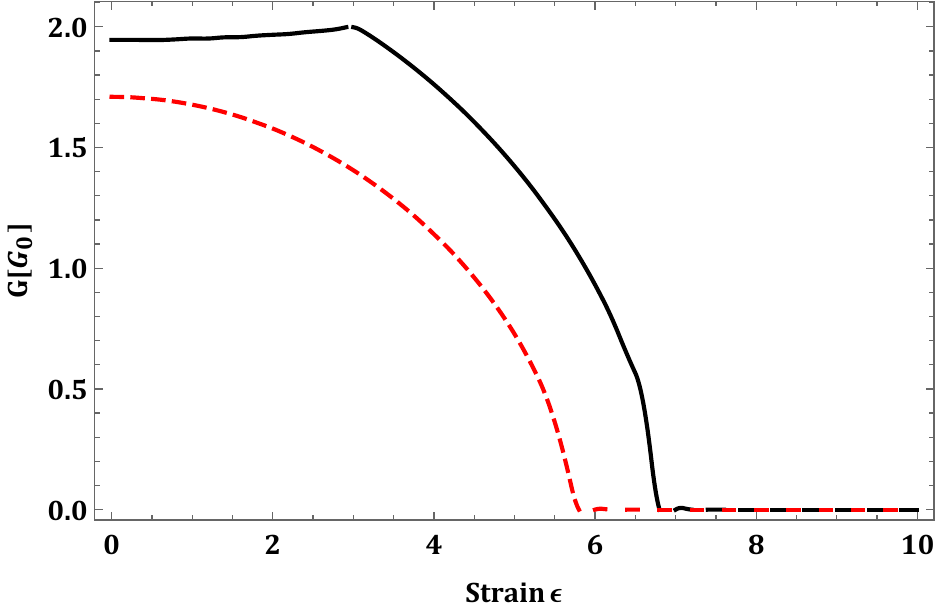}}\label{phic}
	\caption{Conductance as a function of strain $\varepsilon$, for $D=5$ nm and $V_0=3$ eV. Spin up (black line) and spin down (red line).}\label{Conduct}
\end{figure}

Figure~\ref{Conduct} displays the relationship between conductance and strain at different incident energy levels. The conductance exhibits oscillations that directly mirror the strain-dependent transmission resonances, a well-known signature of coherent transport through a finite barrier \cite{Datta1995,Buttiker1986}. 
{At zero strain ($\varepsilon=0$), the conductance corresponds to the unperturbed junction and is determined solely by the electrostatic barrier geometry, the incident energy, and the intrinsic band structure of monolayer WSe$_2$. In this case, transport is governed by the available propagating modes, leading to a baseline conductance that is independent of mechanical deformation and shows no strain-induced modulation. When uniaxial strain is applied, the band structure is modified through a valley-dependent shift in the electronic dispersion, which changes the phase accumulation and transmission resonances inside the barrier. As a result, the conductance deviates from its zero-strain baseline and develops well-defined oscillations driven by strain-controlled interference and mode-matching effects.} 
At lower energies, only a limited number of transverse modes contribute to transport, resulting in pronounced oscillations. As energy levels increase, additional modes become accessible which create a smoother conductance pattern through the combination of angular distribution and multiple modes, according to the standard Landauer--Büttiker theory of conductance quantization through mode summation \cite{Buttiker1986}. The weak-strain regime shows conductance that maintains almost the same value because small lattice deformation affects both band dispersion and the number of active propagating channels only in minor ways, which corresponds to the strain-induced band structure findings for $2$D materials \cite{Roldan2015,CastroNeto2009}. The experiment demonstrates a distinct division between spin-up and spin-down conductance, which reflects the strain-driven development of spin polarization. The results demonstrate that strain changes both the strength of conductance and the ability to manage spin-specific transport pathways. Stress-driven changes in spin--orbit interaction and band edge positions create spin-dependent separation through their effect on spin degeneracy, resulting in distinct transmission patterns for different spin channels as shown in spin--orbit-coupled monolayers and transition metal dichalcogenides \cite{Roldan2015,Xiao2012}. The conductance exhibits a continuous decline as strain increases at high energy levels, indicating that propagation modes become less active due to strain-induced modifications of the band structure. These results suggest that mechanical deformation in WSe$_2$ controls conductance, providing a way to design tunable 2D nanoelectronic and spintronic devices.

\section{Polarization}\label{pol}
From the conductance, one can calculate the spin and valley polarization to better understand the influence of the proposed structure on these properties;   This provides an additional way to exploit WSe$_2$ in  spintronics and valleytronics applications. The polarizations are defined as:
\begin{align}
	& 
P_s = \frac{G_\tau^{(\uparrow)} - G_\tau^{(\downarrow)}}{G_\tau^{(\uparrow)} + G_\tau^{(\downarrow)}}
\\
&P_v=\frac{G_\tau-G_{-\tau}}{G_\tau+G_{-\tau}}
\end{align}
The quantities $G_\tau^{(\uparrow)}$ and $G_\tau^{(\downarrow)}$ denote the conductances for spin-up and spin-down electrons in valley $\tau$, while $G_\tau$ and $G_{-\tau}$ represent the conductances in the two valleys.
%\section{Numerical results}

\begin{figure}[ht]
	\centering
	\subfloat[ Valley {$K$}]{\includegraphics[scale=0.27]{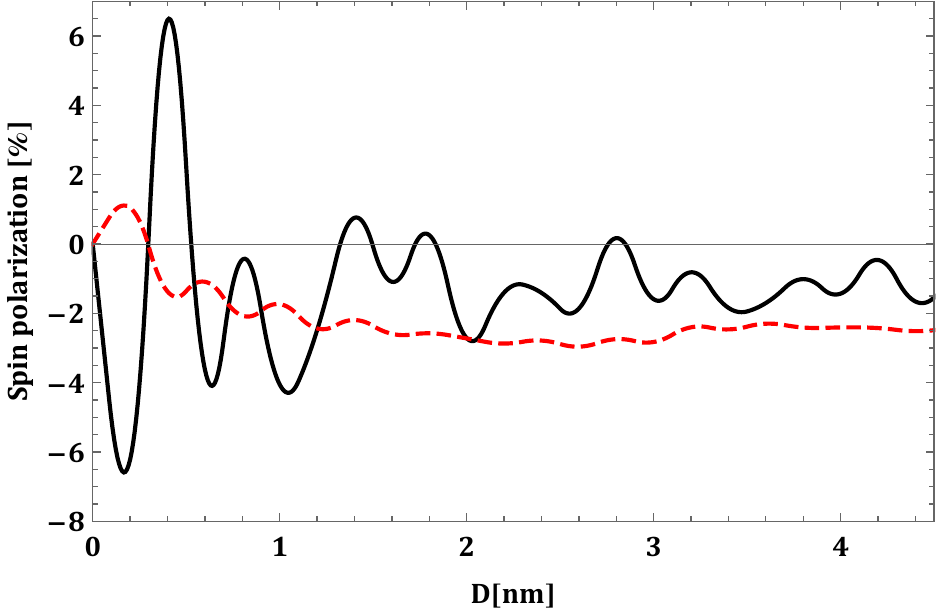}}\label{phia}\subfloat[Valley {$K^\prime$} ]{\includegraphics[scale=0.27]{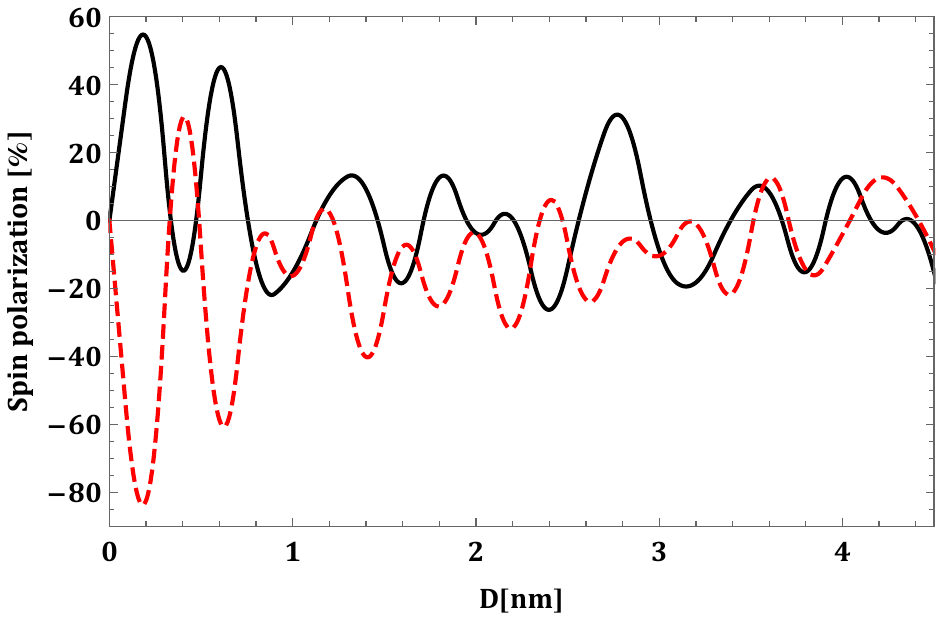}}\label{phib}
	\subfloat[Spin up]{\includegraphics[scale=0.27]{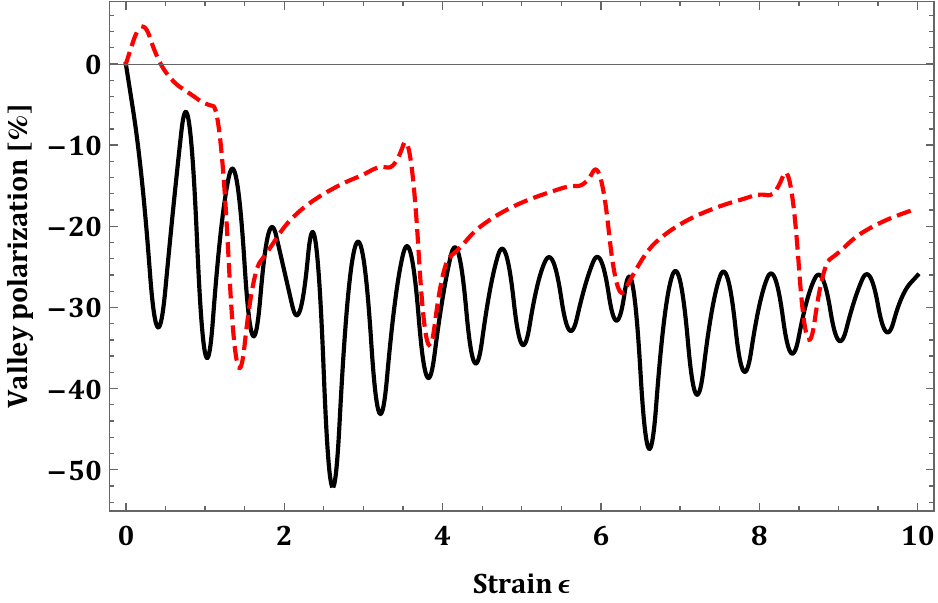}}\label{phic}\subfloat[Spin down]{\includegraphics[scale=0.27]{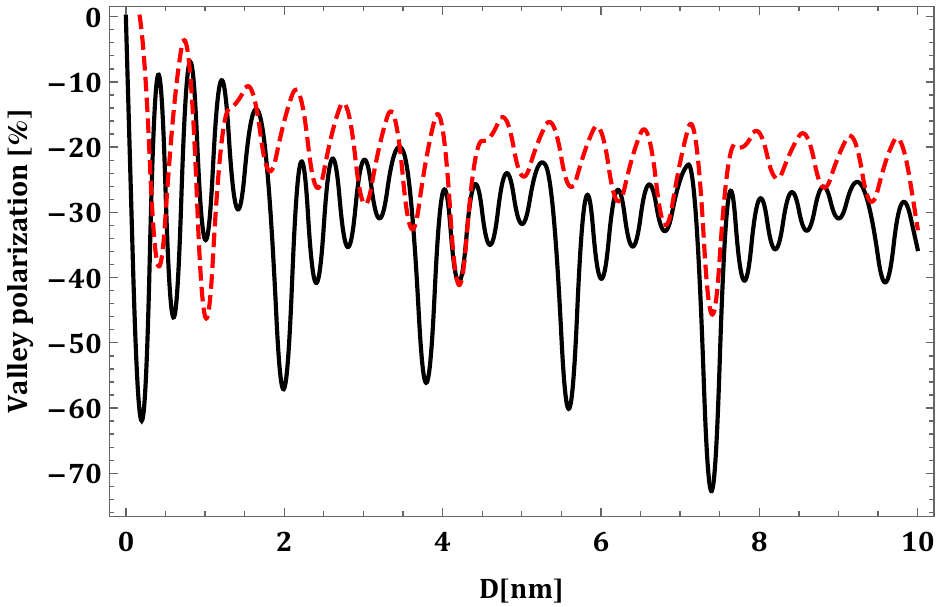}}\label{phic}
	\caption{Polarization as a function of barrier width $D$ for $\varepsilon=4$, $V_0=3$ eV, $E=1.1$ eV (black line) and $E=1.3$ eV (red dashed line).}\label{Pol}
\end{figure}

Figure~\ref{Pol} shows how both spin and valley polarization change when the barrier width $D$ varies. The two quantities display periodic patterns because their movement derives from two different types of spin and valley-based modes that interact inside the barrier. The multiple reflections at the barrier interfaces create oscillations, which produce Fabry-Pérot resonances that respond to spin-orbit coupling and valley-dependent band dispersion in single-layer transition metal dichalcogenides \cite{Kormanyos2015kp, Xiao2012}. 
{To clarify the origin of the polarization, the observed oscillatory behavior with barrier width originates from Fabry–Pérot interference inside the barrier region. The spin–orbit coupling induces a spin-dependent phase accumulation, while the finite barrier width controls the constructive and destructive interference conditions for different spin and valley channels.} 
The $K$ and $K'$ valleys show different spin polarization patterns because WSe$_2$ possesses an inherent relationship between its spin and valley degree of freedom properties. 
{Moreover, strain modifies the valley-dependent electronic band structure and changes the effective propagation wave vectors inside the barrier. This strain-induced asymmetry alters the constructive and destructive interference conditions differently for the $K$ and $K'$ valleys, thereby enhancing the valley polarization and enabling selective control of valley transport.} 
The polarization magnitude reaches its maximum value at minimal barrier widths because this creates maximum confinement and carriers maintain their phase coherence properties. The barrier functions as a resonator that selectively amplifies spin and valley signals, which leads to increased polarization through constructive interference according to theoretical research about spin and valley transport in two-dimensional semiconductors \cite{Zarenia2013,Wang2015}. 
{The barrier width $D$ therefore governs the phase coherence and resonance conditions inside the structure. For small values of $D$, the coherent transport is more pronounced and constructive interference is enhanced, resulting in larger spin and valley polarization amplitudes.} 
The oscillation patterns decrease when $D$ increases because coherent interference becomes less effective over extended distances. The valley polarization demonstrates a major difference between spin-up and spin-down channels because the combination of strain and barrier engineering enables researchers to control both spin and valley components separately. The current results demonstrate that barrier geometry and strain interact to control spin and valley polarization in WSe\(_2\)-based structures, which differs from earlier theoretical research findings about this topic \cite{NanoEnergy2020,PhysicaE2024,Yang2025}.

\begin{figure}[ht]
	\centering
	\subfloat[${K}$ (black), ${K'}$ (red)]{\includegraphics[scale=0.27]{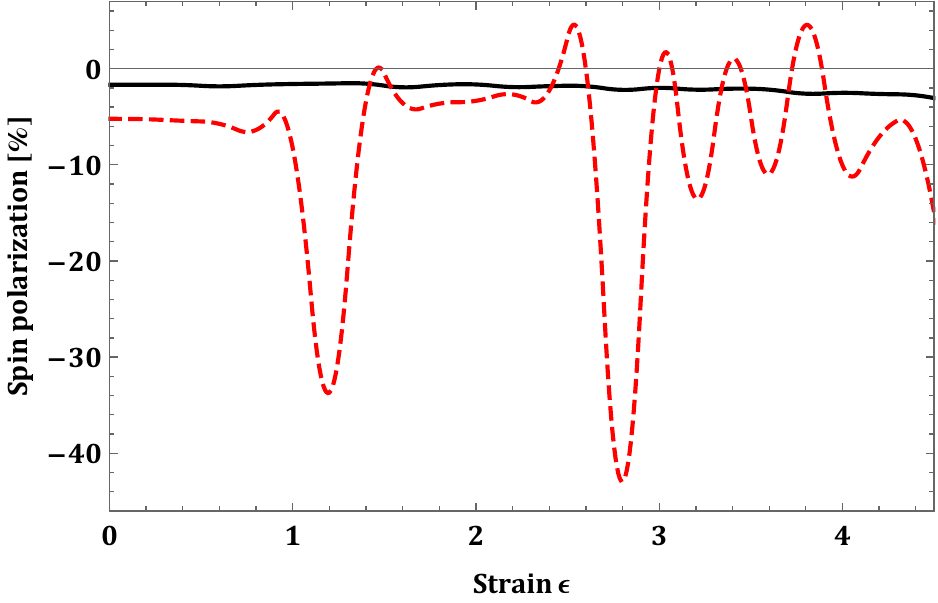}}\label{polstraina}\subfloat[Spin up (black), down (red)]{\includegraphics[scale=0.27]{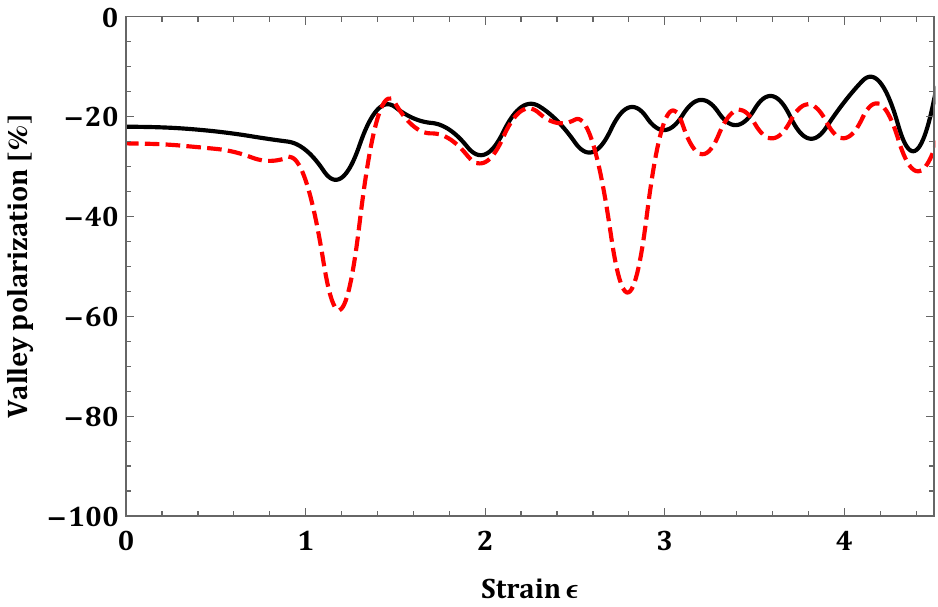}}\label{polstrainb}
	\caption{Polarization as a function of strain $\varepsilon$ for $D=5$ nm, $V_0=3$ eV and $E=1.3$ eV.}\label{polstrain}
\end{figure}

The uniaxial strain $\varepsilon$ affects spin and valley polarization according to Figure~\ref{polstrain} which analyzes a barrier width of $D = 5$~nm and a barrier height of $V_0 = 3$~eV and an incident energy of $E = 0.6$~eV. The spin polarization of Fig.~\ref{polstrain}a shows valley-specific results with the black curve representing the $K$ valley and the red dashed curve representing the $K'$ valley. 
{In the absence of strain ($\varepsilon=0$), both spin and valley polarizations are nearly zero, reflecting symmetric transport between spin-up and spin-down channels and between the $K$ and $K'$ valleys. In this regime, no external symmetry-breaking mechanism is present. When strain is applied, a valley-dependent gauge field is generated, which modifies the effective longitudinal momentum differently for the $K$ and $K'$ valleys.} 
The low strain conditions show that both valleys maintain spin polarization near zero because carriers can transmit equally in spin-up and spin-down states. The spin degeneracy exhibits weak reduction because strain-induced gauge fields show minimal strength \cite{CastroNeto2009,Guinea2010}. The $K'$ valley displays sharp negative spin polarization dips which achieve $-40\%$ at approximate strain levels of $\varepsilon \approx 1$ and $\varepsilon \approx 2.9$.  The $K$ valley shows only minor fluctuations in spin polarization which continue throughout the entire strain period. 
{The origin of this valley-dependent spin polarization is the combined action of strain and intrinsic spin--orbit coupling. While the spin--orbit coupling lifts the spin degeneracy and produces spin-dependent transmission channels, the applied strain modifies the resonance conditions differently in the two valleys through the strain-induced gauge potential.} 
The valley asymmetry results from strain-induced gauge fields which change effective longitudinal momentum differently for both valleys. The $K'$ valley allows one spin species to meet resonant tunneling conditions while the $K$ valley shows reduced tunneling as coherent mesoscopic transport pattern through barriers predicts \cite{Datta1995,Young2009}.
The presence of extremely narrow polarization minima shows that spin filtering operates with high selectivity at strain conditions which cause destructive interference to nearly eliminate one spin pathway. The strain range beyond critical level shows oscillations which result from continuous spin-dependent resonance points crossing through the barrier. 
{In addition, the barrier parameters, particularly the barrier width and height, determine the phase accumulation and the Fabry–Pérot resonance conditions inside the barrier region. These parameters therefore control the positions and amplitudes of the polarization oscillations.} 
The oscillations decrease their strength because the higher strain levels cause coherent interference effects to become less effective. Valley polarization which shows the spin-up state as a black curve and the spin-down state as a red dashed curve appears in Fig.~\ref{polstrain}b. The valley polarization in low-strain conditions shows a negative value which stays almost unchanged because one valley (here $K'$) dominates both spin channels. The valley polarization reaches extreme negative values after strong oscillations begin to develop when strain increases with the spin-down channel reaching values near $-60\%$ at  $\varepsilon \simeq 1.1$ and again at  $\varepsilon \simeq 2.9$. The pronounced minima indicate that strain more effectively improves the selectivity of the valley for this spin orientation. The spin-up channel shows oscillatory patterns which display smaller amplitude than the main oscillations because the strain effect interacts with spin--orbit coupling to create valley splitting.  The phase shift between the spin-up and spin-down curves shows that the transmission thresholds for different spin species experience different effects from strain. The strain-controlled Fabry-Pérot interference produces oscillations which occur inside the barrier because the gauge field that depends on strain governs the phase differences between the two valleys \cite{Young2009}. The selective control of valley-resolved transport through spin orientation shows that strain can either enhance or suppress specific transport modes. The results show that uniaxial strain functions as a powerful external control mechanism which enables simultaneous spin and valley filtering in the WSe$_2$ monolayer barrier structure according to \cite{Xu2014,Kosmider2013}.

\section{Physical picture} \label{phy}
Figures~\ref{phi}--\ref{polstrain} collectively demonstrate that uniaxial strain, barrier geometry, and electrostatic potential provide a highly tunable framework for controlling tunneling, conductance, and spin--valley polarization in monolayer WSe$_2$. The transmission exhibits robust Klein tunneling at normal incidence despite the finite band gap, while angular dependence and barrier width introduce Fabry-Pérot interference patterns due to multiple reflections inside the barrier. 

{
	To clarify the specific role of strain, it is useful to compare the strained and unstrained transport regimes. In the absence of strain (\(\varepsilon=0\)), the transport properties are governed solely by the electrostatic barrier geometry and the intrinsic electronic structure of monolayer WSe$_2$. In this unstrained regime, the transmission and conductance resonances originate only from conventional Fabry-Pérot interference inside the barrier, while spin and valley polarizations remain weak because transport is nearly symmetric between valleys and spin channels. When uniaxial strain is applied, it acts as an effective valley-dependent gauge field that shifts the longitudinal momentum and modifies the resonance conditions inside the barrier. Hence, the phase accumulation becomes strain-dependent, leading to oscillatory transmission and conductance as functions of strain, barrier height, and energy. These strain-controlled resonances produce strong modulations in spin- and valley-resolved conductance together with enhanced spin- and valley-selective transport. Therefore, the pronounced modulation of resonances and the emergence of strong spin and valley filtering are direct consequences of the applied strain.
}
As a consequence, large and tunable spin and valley polarizations emerge, exhibiting notable asymmetry between the $K$ and $K'$ valleys and between spin-up and spin-down channels. In particular, strain enables selective enhancement or suppression of specific spin--valley transport channels, allowing the same junction to operate as a reconfigurable spin and valley filter. Altogether, these results establish strained WSe$_2$ barriers as a versatile platform for engineering coherent spin- and valley-dependent quantum transport, with promising implications for valleytronic and spintronic device applications based on two-dimensional transition-metal dichalcogenides \cite{Mak2010,Rycerz2007}.

{The oscillatory features observed in transmission and conductance can be attributed to quantum interference effects between phase-coherent electron wave functions propagating through the barrier region. The applied uniaxial strain modifies the electronic band structure and induces a strain-dependent phase accumulation along the transport direction. As a result, electron waves undergo constructive and destructive interference depending on strain, energy, barrier width, and incident angle, leading to the observed oscillatory patterns. These features can also be interpreted as Fabry-Pérot-like resonances arising from multiple reflections inside the barrier region.  A similar compression-induced quantum interference effect has been reported in chiral molecular systems \cite{Wang2026_PRB113_104416}, where mechanical deformation modifies the phase coherence of electronic transport. However, the underlying mechanism in the present study is distinct, since the phase modulation arises from strain-induced changes in the Dirac-like band structure of a two-dimensional semiconductor rather than from molecular chirality. Nevertheless, both systems share the same fundamental principle of phase-coherent quantum interference governing transport oscillations.
}

\section{Conclusion}\label{con}
We have theoretically studied electronic transport in tungsten diselenide WSe$_2$ monolayer under uniaxial strain and an external scalar potential. The system is modeled using an electrostatic barrier that divides the monolayer into three regions. After solving the Dirac equation, we have calculated the eigenvalues and associated eigenvectors in each region.  Applying the continuity conditions at the interfaces, we have obtained the transmission  probability, which has then been used to compute the conductance, the spin and valley polarizations depending on the physical parameters of our system. 

Subsequently, the results showed that the electronic transport properties of WSe$_2$ strongly depended on the the uniaxial strain and barrier parameters. Indeed, the transmission reaches an almost unity value at normal incidence, confirming the persistence of Klein tunneling despite the presence of an energy gap induced by spin–orbit coupling. In contrast,  as the incidence angle increases, the transmission gradually decreases and vanishes beyond a critical angle, determined by the incident energy and the strain intensity, due to the conservation of transverse momentum. Furthermore, variations in the barrier width and barrier height lead to pronounced oscillations in the transmission and conductance, which are signatures of Fabry–Pérot–type quantum interference. In this regime, large and tunable spin- and valley-polarized transport is also observed, resulting from the selective separation of transport channels. In particular, a strong asymmetry between valleys $K$ and $K'$ appears under the effect of uniaxial strain, allowing effective and independent control of spin and valley degrees of freedom through barrier tuning. 

Notably, the combination of mechanical deformation and scalar potential enables precise tuning of transmission, conductance, spin and valley polarization. Experimental studies  \cite{Zhu2011,Wang2015_WSe2,Xu2014_WSe2} demonstrate that these effects can be  realized in practical platforms. This high degree of tunability makes strain-engineered barriers in WSe$_2$ an attractive framework for manipulating coherent quantum transport. Such controllability opens promising perspectives for spintronic, valleytronic, nanoelectronic, and optoelectronic applications based on two-dimensional materials.

\end{document}